\documentclass[doublecol]{epl2}
\usepackage{graphicx}

\title{Doping-dependent Phase Diagram of LaO{\it M}As ({\it M}=V--Cu)
and Electron-type Superconductivity near Ferromagnetic Instability}

\author{Gang Xu$^1$, Wenmei Ming$^1$, Yugui Yao$^1$, Xi Dai$^1$,
Shou-Cheng Zhang$^2$, and Zhong Fang$^1$}
\shortauthor{Gang Xu\etal}

\institute{
  \inst{1}Beijing National Laboratory for Condensed Matter
Physics, and Institute of Physics, Chinese Academy of Sciences,
Beijing 100190, China\\
  \inst{2}Department of Physics, Stanford University,
Stanford, CA 94305-4045 }


\abstract{By first-principles calculations, we present a
doping-dependent phase diagram of LaO{\it M}As ({\it M}=V--Cu)
family. It is characterized as antiferromagnetic semiconductor
around LaOMnAs side and ferromagnetic metal around LaOCoAs. Both
LaOFeAs and LaONiAs, where superconductivity were discovered, are
located at the borderline of magnetic phases. Extensive Fermi
surface analysis suggests that the observed superconductivity is of
electron-type in its origin. We discuss possible pairing mechanisms
in the context of competing ferromagnetic phases found in this work
and the ferromagnetic spin fluctuations.}

\begin{document}

\maketitle

The studies on new superconductors, particularly non-cuperate
layered compounds, are always exciting and open up new windows for
the possible further raising of transition temperature $T_c$.
Except few of non-transition-metal compounds, such as MgB$_2$, where
superconductivity with $T_c$ up to 39K was found~\cite{MgB2},
problems in many cases are as complicated and challenging as what we
found in cuperates.  In the layered ruthenates, the ferromagnetic
spin fluctuation is important and spin-triplet $p$-wave character
was suggested~\cite{SRO1} for the superconductivity found in
Sr$_2$RuO$_4$ with $T_c\sim$1K~\cite{SRO2}. In a more recent example
Na$_x$CoO$_2$$\cdot$(H$_2$O)$_y$~\cite{NCO1}, the geometry
fluctuations due to triangle-lattice are extensively
discussed~\cite{NCO2}. Here we will show that a rich
doping-dependent phase diagram can be realized in the new family of
layered compounds LaO{\it M}P or LaO{\it M}As ({\it M}=V--Cu), where
up to $T_c$=26K superconductivity was reached very recently in
LaOFeAs after F$^-$ doping~\cite{LOFS1,LOFS2,LOFS3}.

The quaternary oxypnictides LaO{\it M}As crystallize in layered
tetragonal structure with $P4/nmm$ symmetry~\cite{LOFS0}. Each
transition-metal (oxygen) layer is sandwiched by two
nearest-neighbor As (La) atomic layers, which form edge-shared
tetrahedrons around the {\it M} (oxygen) sites. The ({\it M}As)$^-$
and (LaO)$^+$ triple-layer-subgroups stack alternatively along the
$c$-axis. The positions of La or As sheets are determined by two
internal parameters, $z_{La}$ and $z_{As}$, which define the
inter-layer distances of La-O and {\it M}-As, respectively. It is
important that this series of compounds are chemically stable such
that systematical tuning is available without altering the structure
and symmetry significantly. For instance, a variety of compounds can
be synthesized by the replacement of transition-metal elements,
where both the electron doping and hole doping can be realized by
replacing O$^{2-}$ or La$^{3+}$ ions. Except the early report for
the structure study~\cite{LOFS0}, the detailed studies on the
electronic and magnetic properties for this series of compounds are
still in its early stage. It was first reported in 2006 that
superconductivity can be realized in LaOFeP below 4K, and the $T_c$
was increased to 7K by F-doping~\cite{LOFS1}. Later,
superconductivity with $T_c$ about 2K was reported for
LaONiP~\cite{LOFS2}, and $T_c$ around 26K was reached very recently
in LaOFeAs again after F-doping~\cite{LOFS3}. We will present in
this letter that both {\it M}=Fe and Ni compounds locate at special
positions of the global phase diagram for the series of {\it M}
substituted compounds. The competing magnetic and superconducting
phases found in our global phase diagram provide important clues on
the possible pairing mechanism in this class of materials.

The phase diagram is constructed from first-principles calculations
based on density functional theory with generalized gradient
approximation (GGA) of PBE-type~\cite{PBE} for the
exchange-correlation potential. We use the plane-wave pseudopotential
method, and the ultra-soft pseudopententail scheme~\cite{Ultra} is
adopted. The convergence of total energy calculations with respect to
number of K-points and cut-off energy (of plane wave expansion) is
well checked, and final results are double checked using the
full-potential linearized augmented plane-wave (FLAPW) method (WIN2K
package)~\cite{WIN2K}. The series of LaO{\it M}As compounds with {\it
M} ranging from V to Cu are all studied with full lattice
optimizations using experimental 2 Fe cell, and the non-magnetic (NM),
ferromagnetic (FM), and (intra-layer) antiferromagnetic states are
treated. The same approach has been also applied to LaO{\it M}P
series, qualitatively same results are obtained, we therefore
concentrate our following discussions on LaO{\it M}As series for
consistence.

As shown in Table I, the optimized lattice parameters and internal
coordinates for Fe and Ni compounds are in excellent agreement with
available experimental data~\cite{LOFS1,LOFS2,LOFS3,LOFS0}, which
demonstrates the quality of our present calculations.  The optimized
parameters are used in our calculations for all the compounds. Moving
from V to Cu, the lattice parameters are only slightly modified,
despite of the dramatical change of number of $3d$ electrons (from
$d^3$ of V$^{2+}$ to $d^9$ of Cu$^{2+}$), suggesting the less
sensitivity of lattice distortion.

\begin{table}
\caption{The optimized lattice parameters ($a$ and $c$), internal
coordinates ($z_{La}$ and $z_{As}$), and calculated specific heat
coefficient $\gamma_0$ and bare susceptibility $\chi_0$ for series
of compounds LaO{\it M}As in the NM state.}

\scalebox{0.75}{
\begin{tabular}{l|c|c|c|c|c|c}
\hline
     &$a$(\AA)  &$c$(\AA)   &$z_{La}$ &$z_{As}$  &$\gamma_0$($\frac{mJ}{k^2mol}$)
   &$\chi_0$($10^{-5}\frac{emu}{mol}$)             \\ \hline
V   &3.9965   &9.3655   &0.1367    &0.1643   &7.47   &10.25    \\
Cr  &3.9355   &9.3467   &0.1397    &0.1592   &9.06   &12.43    \\
Mn  &4.0355   &8.7822   &0.1438    &0.1427   &11.37   &15.59    \\
Fe  &4.0396   &8.6289   &0.1461    &0.1369   &5.52   &7.57    \\
Co  &4.0621   &8.5434   &0.1466    &0.1369   &7.41   &10.17    \\
Ni  &4.1404   &8.3139   &0.1466    &0.1366   &3.81   &5.23    \\
Cu  &4.1442   &8.5681   &0.1424    &0.1539   &3.98   &5.46    \\
\hline
\end{tabular}
}
\end{table}

Fig.1 shows the phase diagram computed for the whole range of
compounds. The solid lines and the dashed lines represent the
stabilization energies of FM and AF states relative to NM solution,
respectively. Two distinct phase regions can be identified: the dome
below Fe gives AF ground state, while the one around Co is
ferromagnetic. The computed magnetic moments (see Fig.1) show that the
left-hand side AF phase region are strongly spin polarized, while the
magnetic moment of the ferromagnetic phase around Co is small,
suggesting possibly different origins of two magnetic phase
regions. It is important to note that both LaOFeAs and LaONiAs are
located at special positions of the phase diagram: Fe compound is at
the borderline between the AF and FM phase regions; while Ni compound
is at the other border of the FM phase. The overall phase diagram
clearly suggests that the magnetic instabilities are crucially
important to understand the physical properties of this new family of
compounds.  The present calcualtions are done based on experimental
crystal cell with 2 Fe per cell~\cite{LOFS3}. However, we notice that
strong nesting effect exists particularly for LaOFeAs. This will lead
to stripe-type spin-density-wave (SDW) ground state with
$\sqrt{2}\times\sqrt{2}$ super-cell structure.  The detailed results
for this SDW state will be presented in a separate paper combined with
experimental results~\cite{SDW}.

\begin{figure}
\includegraphics[clip,scale=0.45]{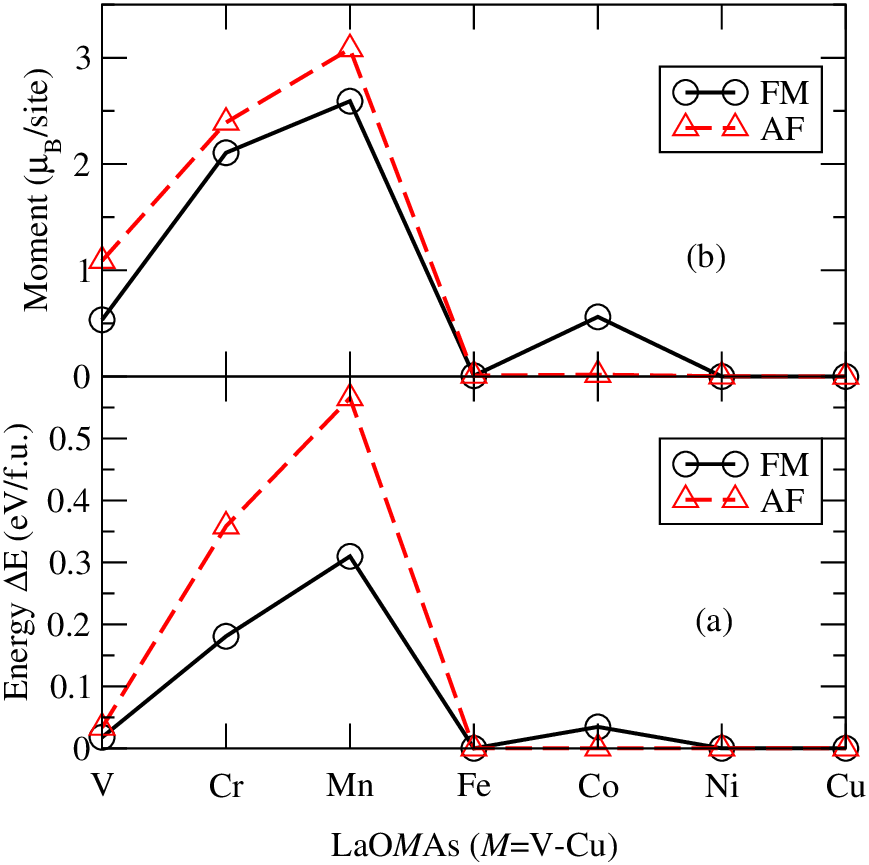}
\caption{Phase diagram: (a) The calculated stabilization energies of
FM and AF states with respect to NM state for various compounds with
increasing number of $3d$ electrons. (b) The corresponding magnetic
moments obtained from calculations.}
\end{figure}

To understand the electronic structures, we show the total and
projected density of states (DOS) of various compounds in Fig.2.
Let's start from LaOFeAs (Fig.2(a)). The states between -2eV and
+2eV are mostly from Fe-$3d$ states, just below which are the states
of O-$p$ and As-$p$ (from -6eV to -2eV). The $p$-$d$ hybridization
between O and Fe is negligible, while that between As and Fe is
sizable. This As-{\it M} $p$-$d$ hybridization is enhanced by
changing to Ni compound (see the projected DOS of Fig.2(b)). Since
the transition metal sites are coordinates by As-tetrahedron, the
crystal field will normally splits the five $d$ orbitals into
low-lying two-folds $e_g$ states and up-lying three-folds $t_{2g}$
states. However, the As-tetrahedron are actually much distorted from
its normal shape (squeezed along $c$ by about 20\%). This distortion
will further split the $e_g$ and $t_{2g}$ manifolds significantly
making the final orbital distributions complicated. As the results,
what actually happens is opposite to what we expect from simple
tetrahedra crystal field: the low-lying manifold is three folds and
the higher-lying manifold is two folds, between which a pseudo-gap
about 0.5eV exists. For Fe compound, the nominal number of $d$
electrons is 6, and the low-lying three-fold bands are nearly fully
occupied with the Fermi level $E_f$ located very close to the deep
of the pseudo-gap. If the crystal field is strong enough, which may
be achieved by the substitution of As atoms, a simple band-insulator
will be expected by enlarging the pseudo-gap.  From the calculated
DOS at $E_f$, we estimate the bare susceptibility and specific heat
coefficient, which are $\chi_0$=7.6$\times10^{-5}$ emu/mol and
$\gamma_0$=5.5 mJ/K$^2$mol for Fe compound (see Table I for other
compound). Taking the experimental susceptibility about
50$\times10^{-5}$ emu/mol at 300K, a renormalization factor about
6.6 is suggested, which is close to that shown in
Sr$_2$RuO$_4$~\cite{SRO3}.

Moving to Mn compound, the half filled $d$-shell (about $d^5$) will
gain energy from the Hund's coupling, and the spin polarized states
will be favored as the results. In reality, the calculated total
energies of both FM and AF solutions are much lower than that of NM
solution (about 0.3eV/f.u for the former and 0.55eV for the later).
The reason why the AF solution is more favored is that a gap about
0.2eV is opened in the AF solution (see Fig.2(e)). It is mentioned
(but without showing data) in the recent experimental
paper~\cite{LOFS1} that LaOMnP is a semiconductor. This is
consistent with our prediction, and further we show that the ground
state of Mn compound is AF. The calculated spin moment is about
3.1$\mu_B$/Mn, which is much reduced from the expected 5$\mu_B$/Mn
of high spin state. There are two possibilities to explain the
reduced moment. One is that Mn is in intermediate spin state rather
than high-spin state. If this is the case, certain kinds of orbital
ordering would be expected for an AF insulator. However, our
calculated occupation numbers of projected $3d$-shell orbitals are
quite uniform, suggesting this possibility is unlikely.  The second
possibility is due to either $p$-$d$ or $d$-$d$ hybridization
(particularly for narrow gap system). The calculated spin moment is
about 4.3$\mu_B$/Mn for typical high-spin AF insulator
MnO~\cite{MnO}, where only the $p$-$d$ hybridization is important.
However, here we point out that the Mn-Mn distance in LaOMnAs is
about 2.8\AA, which is much shorter than what was found in MnO
(about 3.2\AA), and is actually very close to the distance in
elementary-Mn (about 2.6$\sim$2.7\AA). The direct $d$-$d$
hybridization will be much enhanced by such short distance, which
will again reduce the moment. The effect of direct $d$-$d$ overlap
has been addressed in previous study for LaOFeP~\cite{LOFP-cal},
however we emphasize here that this is generally true for all the
compounds of this family as shown by the optimized structure (Table
I).

The stabilization of FM phase region at the right hand side has
different origin as will be discussed here for LaOCoAs. Co has one
more $d$ electron than Fe, therefore the Fermi level is lifted up
and located above the pseudo-gap. What is interesting is that a
strong Van-Hove singularity (VHS) is present just at the Fermi level
of LaOCoAS NM DOS as shown in Fig.2(d). The high N($E_f$) in the
presence of this VHS will push the system to be itinerant FM due to
Stoner instability. This mechanism is further supported by the
following factors: (1) the FM region is relatively narrow; (2) the
polarized spin moment is small (about 0.5$\mu_B$/Co); (3) the energy
gain is also small (about 35meV/f.u.). By adding one more electron,
for LaONiAs, the Fermi level is shifted away from the VHS, the
system is recovered to be NM again as shown in the phase diagram.

\begin{figure}
\includegraphics[clip,scale=0.48]{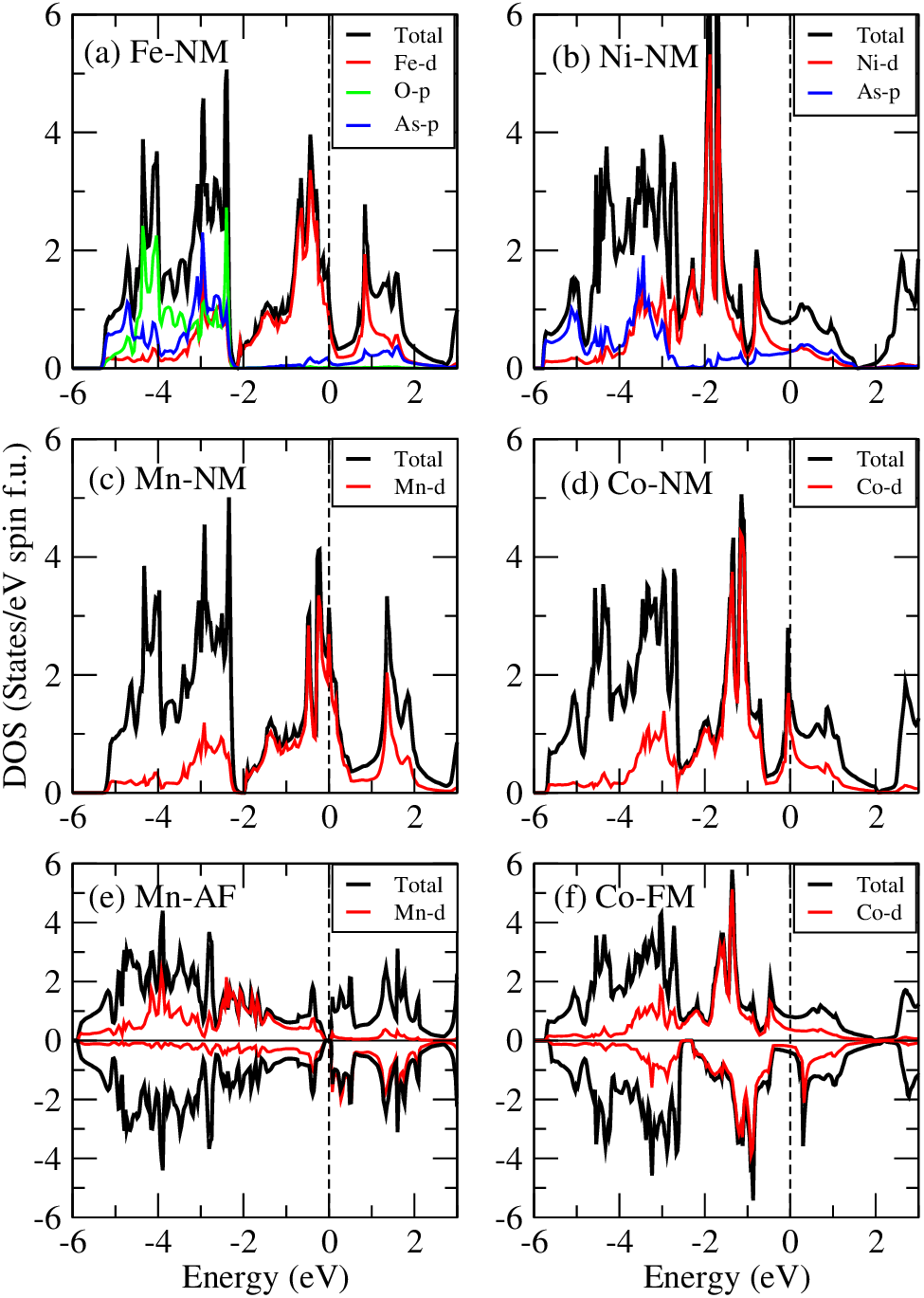}
\caption{The calculated electron density of states (DOS) for various
compounds in different states (see text part for explanation).}
\end{figure}

Having finished the discussions for the phase diagram and the
general picture of the electronic structure, now let us focus on the
LaOFeAs and LaONiAs, where superconductivity are discovered. Fig.3
gives the calculated band structures and Fermi surfaces (FS) for
both compounds (the window for $d$-bands is shown, and there are
totally ten $d$ bands in our unit cell). First of all, the band
dispersions along the $z$ direction are all very weak suggesting the
2-dimensional nature of those compounds. Considering the in-plane
dispersions, the band structure of LaOFeAs can be schematically
separated into two parts. The bands below $E_f$ are relative flat
and have little contribution coming from the As-$p$ states, while
the bands above $E_f$ are quite dispersive (except some flat
branches around +1eV which corresponds to the VHS discussed above),
and have large weight coming from the As-$p$ character (as shown by
the projected fat-bands plot).  Those dispersive bands form
electron-like FS cylinders around the M-A lines of the Brillouin
zone, and hole-like FS cylinders are formed around the $\Gamma$-Z
lines due to the Fermi level crossing of the bands from the lower
part. For LaONiAs, all the flat bands are pushed down to below Fermi
level, and only the dispersive bands remain to cross the Fermi
level, which give the large electron FS around M-A lines, but
hole-type FS around X-R line instead of $\Gamma$-Z.

The experimental results show that the LaOFeAs is superconducting
only after electron-type F-doping~\cite{LOFS3}, and also the $T_c$
of LaOFeP (which has very similar band structure~\cite{LOFP-cal} as
LaOFeAs) is enhanced with F-doping~\cite{LOFS1}. All those evidences
suggest that the electron-like FS formed by the dispersive band
should be responsible for the superconductivity. Taking the
calculated total DOS of LaOFeAs, the $N(E_f)$ will decrease with
up-shifting of Fermi level, in opposite to the experimental trend
observed for F-doping. However, if we only take into account the
contribution from the electron-like FS, its DOS will increase with
increasing Fermi energy. Actually, as shown in Fig.3, hole-like FS
of LaOFeAs after 10\% F-doping are much reduced, and electron-like
FS are enlarged, compared to the case without
doping~\cite{LOFP-cal}. The very recent Hall measurements also
suggested the electron-type conductivity in LaONiAs after
F-doping~\cite{NLWang}.

\begin{figure}
\includegraphics[clip,scale=0.45]{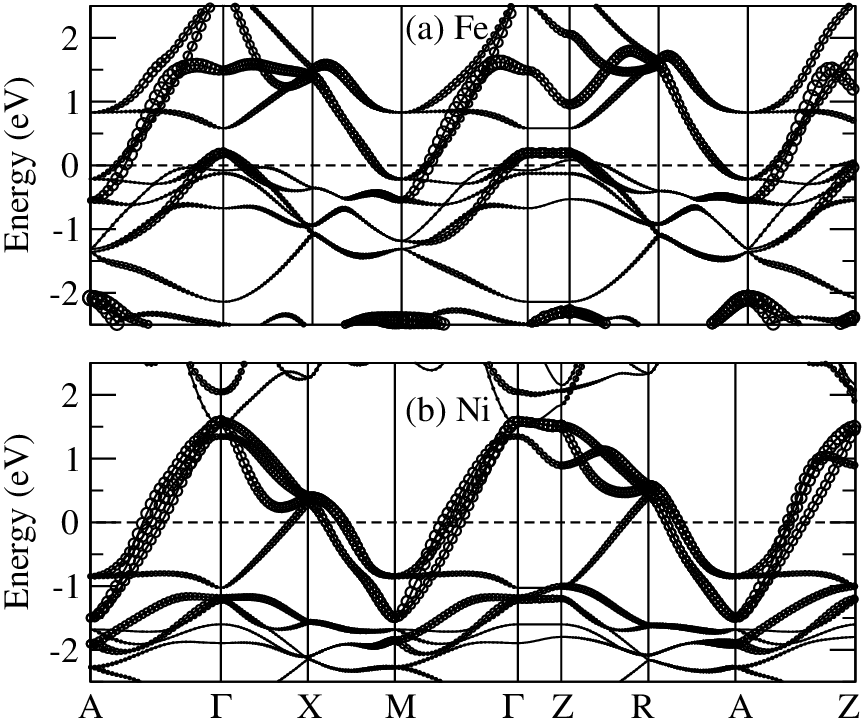}
\includegraphics[clip,scale=0.25]{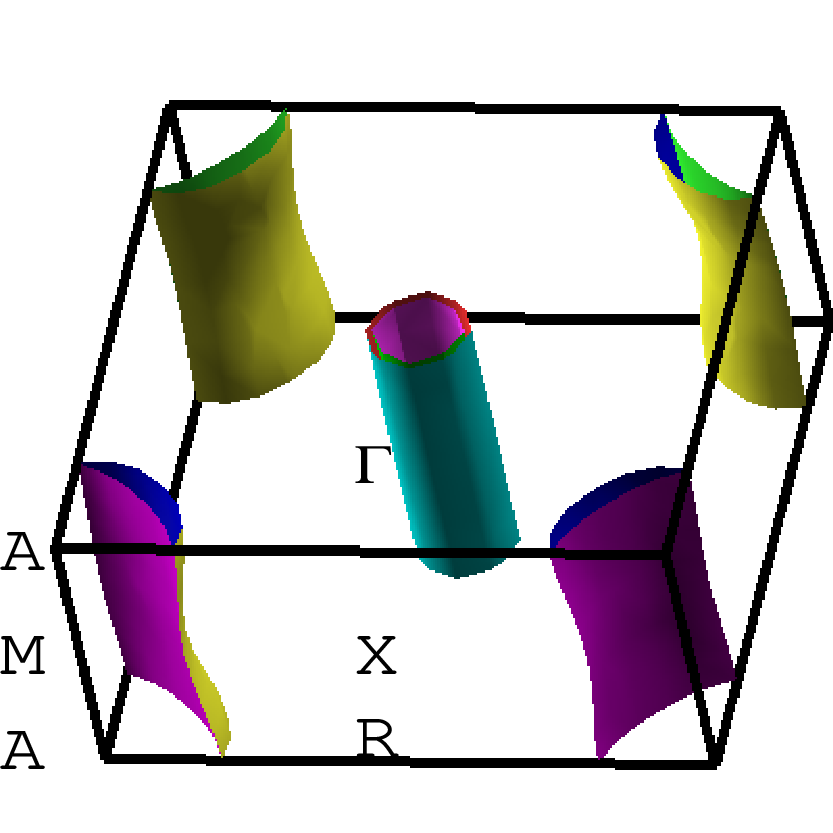} \  \  \  \   \  \
\includegraphics[clip,scale=0.25]{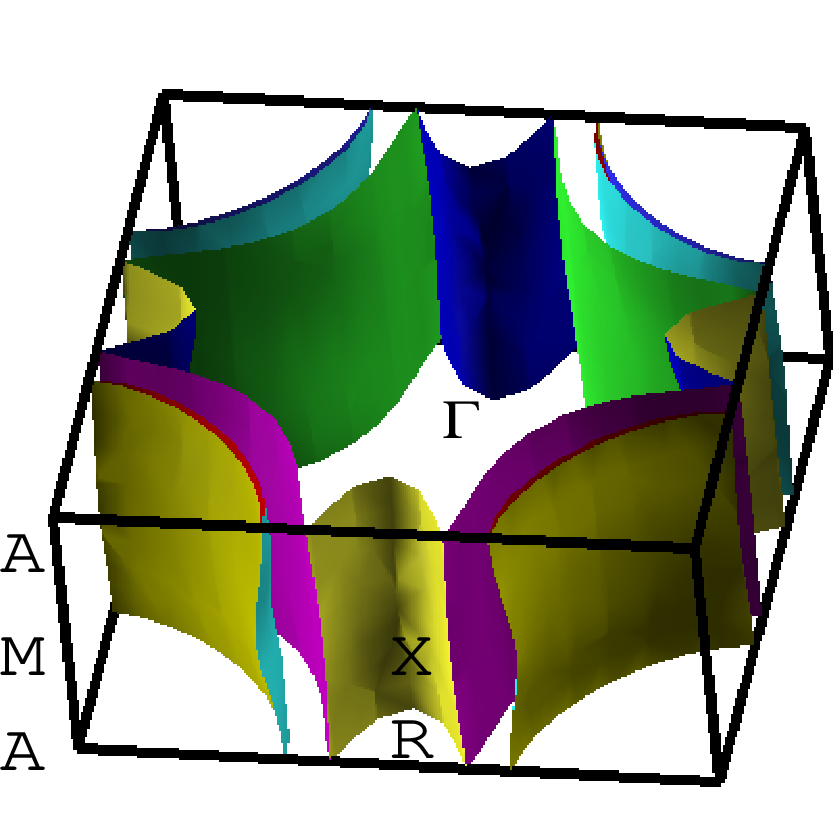}
\caption{The calculated band structures of (a) LaOFeAs and (b)
LaONiAs. The fat-bands are shown with projection to As-$p$ states.
The lower panels show the calculated Fermi surfaces of LaOFeAs after
10\% F-doping (left) and LaONiAs (right).}
\end{figure}

Up to now, superconductivity has been discovered in several transition
metal systems, including the cuperates, ruthenates and
cobaltates. Most of these superconducting materials are hole doped and
only a few of them are electron doped. In cuperates, the electron
doped compounds usually have much lower $T_{c}$ than the hole doped
ones. In sodium cobaltates, only hole-doped compounds are
realized. The present systems are also layered compounds, and the
$p$-$d$ hybridization is sizable. Considering the close vicinity to
the ferromagnetic instability, the present systems are very similar to
Sr$_2$RuO$_4$, where strong ferromagnetic fluctuations favor triplet
pairing. In such case, the question is whether the pairing state is
unitary and time reversal invariant, analogous to the B phase of
$^3$He, or non-unitary and time reversal breaking, analogous to the A
phase of $^3$He~\cite{He3}. Our calculations shows that LaOFeAs can be
basically characterized as low density electron carriers doped on top
of a band insulator with filled $d^6$ valence orbitals. In $^3$He, the
B phase is realized under the low or ambient pressure condition while
A phase is realized under high pressure condition close to
solidification. By analogy, we suggest that the low electron density
system LaOFeAs is in the weak coupling limit, which generally favors
the unitary, or B phase like, pairing symmetry. Upon further
increasing doping, the non-unitary A phase could be realized. In two
dimensions, unitary B state can be characterized as a state where the
up (down) spin electrons are paired in the $p_x+i p_y$ ($p_x-i p_y$)
state, so that the time reversal symmetry is preserved. This state is
similar to the topologically non-trivial state, characterized by $Z_2$
invariant~\cite{Z2}, found in quantized spin-Hall
insulator~\cite{HgTe}. In contrast, the time reversal symmetry
breaking A phase could be realized in Sr$_2$RuO$_4$ partly because the
carriers density is high.

The topological nature of the proposed pairing state for LaOFeAs
implies the existence of counter-propagating edge states which can
be tested experimentally. In the bulk, the pairing state is fully
gapped, and STM experiment would show a full gap in the I-V
characteristics. However, moving to the edge, the STM experiment
would show a gapless spectrum, revealing the gapless edge states
protected by the time reversal symmetry. Most strikingly, in the
presence of a magnetic impurity near the edge, the local density of
states would show a gap again, due to breaking of time reversal
symmetry.

\acknowledgments
We acknowledge the valuable discussions with Y. P.
Wang, N. L. Wang, J. L. Luo, H. H. Wen, T. Hughes, X. L. Qi, S.
Raghu and D. Scalapino, and the supports from NSF of China and that
from the 973 program of China (No.2007CB925000 and 2006CB921300).
SCZ acknowledge supports by the NSF under grant numbers DMR-0342832
and the US DOE, Office of Basic Energy Sciences under contract
DE-AC03-76SF00515.

\end{document}